\documentclass[Physsubmission, Phys]{SciPost}

\hypersetup{
    colorlinks,
    linkcolor={red!50!black},
    citecolor={blue!50!black},
    urlcolor={blue!80!black}
}

\usepackage[bitstream-charter]{mathdesign}
\def\Tr{\text{Tr}} 
\newcommand{\F}{\phantom {1}}
\def\nn{\nonumber}
\newcommand{\beqn}{\begin{eqnarray}}
\newcommand{\beq}{\begin{equation}}
\newcommand{\eeqn}{\end{eqnarray}}
\newcommand{\eeq}{\end{equation}}

\newcommand{\kbra}[1] { \left< #1 \right>}
\usepackage{subcaption}

\DeclareSymbolFont{usualmathcal}{OMS}{cmsy}{m}{n}
\DeclareSymbolFontAlphabet{\mathcal}{usualmathcal}

\begin{document}
\begin{center}{\Large \textbf{
Abelian monopoles of the Dirac type\\
and color confinement in QCD
}}\end{center}

\begin{center}
Tsuneo Suzuki\textsuperscript{1$\star$},
Atsuki Hiraguchi\textsuperscript{2,3} and
Katsuya Ishiguro\textsuperscript{4}
\end{center}

\begin{center}
{\bf 1} RCNP, Osaka University, Ibaraki Osaka 567-0047, Japan\\
{\bf 2} Institute of Physics, National Yang Ming Chiao Tung University
Hsinchu 30010, Taiwan \\  
{\bf 3} Department of Mathematics and Physics, Kochi University,
Kochi 780-8520, Japan\\
{\bf 4} Library and Information Technology, Kochi University,
Kochi 780-8520, Japan\\
* suzuki0256@gmail.com
\end{center}

\begin{center}
\today
\end{center}


\definecolor{palegray}{gray}{0.95}
\begin{center}
\colorbox{palegray}{
  \begin{minipage}{0.95\textwidth}
    \begin{center}
    {\it  XXXIII International (ONLINE) Workshop on High Energy Physics \\"Hard Problems of Hadron Physics:  Non-Perturbative QCD \& Related Quests"}\\
    {\it November 8-12, 2021} \\
    \doi{10.21468/SciPostPhysProc.?}\\
    \end{center}
  \end{minipage}
}
\end{center}

\section*{Abstract}
{\bf
We present results of $SU(3)$ Monte-Carlo studies of a new color confinement scheme due to Abelian-like monopoles of the Dirac type without any  gauge-fixing.  We get (1) perfect Abelian dominance with respect to the static potentials on $12^4\sim 16^4$ lattice at $\beta=5.6-5.8$ using the multilevel method, (2)  monopole as well as Abelian dominances with respect to the static potentials by evaluating Polyakov loop correlators on $24^3\times4$ lattice at $\beta=5.6$. (3) Abelian dual Meissner effects are studied directly by measuring Abelian color fields and monopole currents around the static source. The vacuum in pure $SU(3)$ seems to be of the type 1 near the border between both types, although scaling is not studied yet.}

\section{Introduction}
\label{sec:intro}
Color confinement in  quantum chromodynamics (QCD) 
is still an important unsolved  problem. As a picture of color confinement, 't~Hooft~\cite{tHooft:1975pu} and Mandelstam~\cite{Mandelstam:1974pi} conjectured that the QCD vacuum is a kind of a magnetic superconducting state caused by condensation of magnetic monopoles and  an effect dual to the Meissner effect works to confine color charges. What is the magnetic monopoles in QCD?
 
When we discuss magnetic monopoles as well as electric fields without scalar fields, it seems necessary to introduce some singularities as shown by Dirac~\cite{Dirac:1931} in $U(1)$ quantum electrodynamics. An interesting idea to introduce such a singularity in QCD is to project QCD to the Abelian maximal torus group by a partial (but singular) gauge fixing~\cite{tHooft:1981ht}. In $SU(3)$ QCD, the maximal torus group is  Abelian $U(1)^2$. Then color magnetic monopoles appear as a topological object at the space-like points corresponding to the singulariry of the gauge-fixing matrix. Condensation of the monopoles  causes  the dual Meissner effect with respect to $U(1)^2$. Numerically, an Abelian projection in various gauges such as the maximally Abelian (MA) gauge~\cite{Kronfeld:1987ri,Kronfeld:1987vd} seems to support the conjecture~\cite{Suzuki:1992rw, Chernodub:1997ay}.  
Although numerically interesting, the idea of Abelian projection~\cite{tHooft:1981ht} is theoretically very unsatisfactory. Especially there are infinite ways of such a partial gauge-fixing and whether the 't Hooft scheme is gauge independent or not is not known. 

In 2010  Bonati et al.~\cite{Bonati:2010tz} found an interesting fact that the violation of non-Abelian Bianchi identity (VNABI) exists behind the Abelian projection scenario in various gauges and hence gauge independence is naturally expected. Along this line,   one of the authors (T.S.)~\cite{Suzuki:2014wya} found a more general relation that VNABI is just equal to the violation of Abelian-like Bianchi identities corresponding to the existence of Abelian-like monopoles. A partial gauge-fixing is not necessary at all from the beginning. If the non-Abelian Bianchi identity is broken, Abelian-like monopoles necessarily appear due to a line-like singularity leading to a non-commutability with respect to successive partial derivatives. This is hence an extension of the Dirac idea of monopoles in QED to non-Abelian  QCD. 

\section{Equivalence of VNABI and Abelian-like monopoles}
\label{sec-1}
First of all, we prove that the Jacobi identities of covariant derivatives lead us to conclusion that  violation of the non-Abelian Bianchi identities (VNABI) $J_{\mu}$ is nothing but an Abelian-like monopole $k_{\mu}$ defined by violation of the Abelian-like Bianchi identities without gauge-fixing. 
 Define a covariant derivative operator $D_{\mu}=\partial_{\mu}-igA_{\mu}$. The Jacobi identities are expressed as $\epsilon_{\mu\nu\rho\sigma}[D_{\nu},[D_{\rho},D_{\sigma}]]=0$.
By direct calculations, one gets
\begin{eqnarray*}
[D_{\rho},D_{\sigma}]&=&[\partial_{\rho}-igA_{\rho},\partial_{\sigma}-igA_{\sigma}]\\
&=&-igG_{\rho\sigma}+[\partial_{\rho},\partial_{\sigma}],
\end{eqnarray*}
where the second commutator term of the partial derivative operators can not be discarded, since gauge fields may contain a line singularity. Actually, it is the origin of the violation of the non-Abelian Bianchi identities (VNABI) as shown in the following. The non-Abelian Bianchi identities and the Abelian-like Bianchi identities are, respectively: $D_{\nu}G^{*}_{\mu\nu}=0$ and $\partial_{\nu}f^{*}_{\mu\nu}=0$.
The relation $[D_{\nu},G_{\rho\sigma}]=D_{\nu}G_{\rho\sigma}$ and the Jacobi identities lead us to
\begin{eqnarray}
D_{\nu}G^{*}_{\mu\nu}&=&\frac{1}{2}\epsilon_{\mu\nu\rho\sigma}D_{\nu}G_{\rho\sigma} \nn\\
&=&-\frac{i}{2g}\epsilon_{\mu\nu\rho\sigma}[D_{\nu},[\partial_{\rho},\partial_{\sigma}]]\nn\\
&=&\frac{1}{2}\epsilon_{\mu\nu\rho\sigma}[\partial_{\rho},\partial_{\sigma}]A_{\nu}
=\partial_{\nu}f^{*}_{\mu\nu}, \label{eq-JK}
\end{eqnarray}
where $f_{\mu\nu}$ is defined as $f_{\mu\nu}=\partial_{\mu}A_{\nu}-\partial_{\nu}A_{\mu}=(\partial_{\mu}A^a_{\nu}-\partial_{\nu}A^a_{\mu})\lambda^a/2$. Namely Eq.(\ref{eq-JK}) shows that the violation of the non-Abelian Bianchi identities, if exists,  is equivalent to that of the Abelian-like Bianchi identities.

Let us denote the violation of the non-Abelian Bianchi identities as  $J_{\mu}=D_{\nu}G^*_{\mu \nu}$ and 
 Abelian-like monopoles $k_{\mu}$ without any gauge-fixing as the violation of the Abelian-like Bianchi identities:
\begin{eqnarray}
k_{\mu}=\partial_{\nu}f^*_{\mu\nu}
=\frac{1}{2}\epsilon_{\mu\nu\rho\sigma}\partial_{\nu}f_{\rho\sigma}. \label{ab-mon}
\end{eqnarray}
Eq.(\ref{eq-JK}) shows that $J_{\mu}=k_{\mu}$.

Due to the antisymmetric property of the Abelian-like field strength, we get Abelian-like conservation conditions~\cite{Arafune:1974uy}:
\begin{eqnarray}
\partial_\mu k_\mu=0. \label{A-cons}
\end{eqnarray}

\section{Abelian and monopole static potentials  in $SU(3)$}
\label{sec-2}
In $SU(2)$ QCD, perfect Abelian dominance is proved  without performing any addtional gauge fixing using the multilevel method~\cite{Luscher:2001} in Ref.~\cite{Suzuki:2007jp, Suzuki:2009xy}. Also perfect monopole dominance is proved very beautifully by applying the random gauge transformation as a method of the noise reduction  of measuring gauge-variant quantities in the same reference~\cite{Suzuki:2007jp, Suzuki:2009xy}.  Also in $SU(2)$ QCD, it is shown clearly that the new Abelian lattice monopoles have the continuum limit with the help of the block-spin renormalization group studies of their density~\cite{Suzuki:2017lco} and the effective action~\cite{Suzuki:2017zdh}. 
However,  in $SU(3)$ QCD on lattice,  it is very difficult. First of all, it is not so simple to define Abelian link fields from non-Abelian link fields without any gauge-fixing  in $SU(3)$.  We choose the following method to define the Abelian link field by maximizing the following overlap quantity
\begin{eqnarray}
RA=\sum_{s,\mu}\Tr\left(e^{i\theta_\mu^a(s)\lambda^a}U_\mu^{\dag}(s)\right), \label{RA}
\end{eqnarray}
where $\lambda^a$ is the Gell-Mann matrix and no sum over $a$ is not taken.
This choice in $SU(2)$ leads us to the same Abelian link fields adopted in  Ref.~\cite{Suzuki:2007jp, Suzuki:2009xy}.

For example, we get from the maximization condition of  (\ref{RA}) an Abelian link field $\theta_1(s,\mu)$ corresponding to $\lambda_1$ $(SU(3))$ as 
\begin{eqnarray*}
\theta_1(s,\mu)&=& 
\textrm{tan}^{-1}\frac{Im(U_{12}(s,\mu)+U_{21}(s,\mu))}{Re(U_{11}(s,\mu)+U_{22}(s,\mu))}, \ \ (\textrm{SU3})
\end{eqnarray*}

Once Abelian link variables are fixed, we can extract Abelian, monopole and photon parts from  the Abelian plaquette variable as follows:
\begin{eqnarray}
\theta_{\mu\nu}^a(s) =\bar{\theta}_{\mu\nu}^a(s)+2\pi
n_{\mu\nu}^a(s)\ \ (|\bar{\theta}_{\mu\nu}^a|<\pi),\label{abel+proj}
\end{eqnarray}
where $n_{\mu\nu}^a(s)$ is an integer
corresponding to the number of the Dirac string.
Then an Abelian monopole current is defined by
\begin{eqnarray}
k_{\mu}^a(s)&=& -(1/2)\epsilon_{\mu\alpha\beta\gamma}\partial_{\alpha}
\bar{\theta}_{\beta\gamma}^a(s+\hat\mu) \nonumber\\
&=&(1/2)\epsilon_{\mu\alpha\beta\gamma}\partial_{\alpha}
n_{\beta\gamma}^a(s+\hat\mu)  \label{eq:amon}.
\end{eqnarray}
This definition (\ref{eq:amon})  satisfies the Abelian conservation condition (\ref{A-cons}) and takes an integer value which corresponds to the magnetic charge obeying the Dirac quantization condition~\cite{DeGrand:1980eq}.
 
\subsection{Perfect Abelian dominance in $SU(3)$}
Now let us evlaluate Abelian static potentials through Polyakov loop correlators written by the Abelian plaquette variable defined above:
\begin{eqnarray}
P_{\rm A} = \exp[i\sum_{k=0}^{N_{t}-1}\theta_1(s+k\hat{4},4)] \label{eq-PA}
\end{eqnarray}
Since the above Abelian Polyakov loop operator without any additional gauge-fixing is defined locally, the Poyakov loop correlators can be evaluated through the multilevel method~\cite{Luscher:2001}. Contrary to the $SU(2)$ case in Ref.~\cite{Suzuki:2009xy}, we need much more number of internal updates to get meaningful results. The simulation parameters are shown in Table~\ref{Table1}.
\begin{table}[t]
\caption{\label{SU3multilevel_parameter}
Simulation parameters for the measurement of static potential using multilevel method. $N_{\rm sub}$ is the sublattice size divided and $N_{\rm iup}$ is the number of internal updates in the multilevel method.}
\label{Table1}
\begin{center}
\begin{tabular}{c|c|c|c|c|c}
\hline
$\beta$ &$N_{s}^{3}\times N_{t}$& $a(\beta)$~[fm]& $N_{\rm conf}$ & $N_{\rm sub}$ & $N_{\rm iup}$ \\ 
\hline
5.60 & $12^3 \times 12$ & 0.2235\F & 6 & 2 & 5,000,000\\
5.60 & $16^3 \times 16$ & 0.2235\F & 6 & 2 & 10,000,000\\
5.70 & $12^3 \times 12$ & 0.17016\F & 6 & 2 & 5,000,000\\
5.80 & $12^3 \times 12$ & 0.13642\F & 6 & 3 & 5,000,000\\
\hline
\end{tabular}
\end{center}
\end{table}
\begin{table}[t]
\caption{
Best fitted values of the string tension $\sigma a^2$, the Coulombic coefficient $c$, and the constant $\mu a$ for the potentials $V_{\rm NA}$, $V_{\rm A}$.}
\label{stringtension_multilevel}
\begin{center}
\begin{tabular}{l|c|c|c}
\multicolumn{4}{l}{}\\ 
\hline
$\beta =5.6, 12^3\times 12$& $\sigma a^2$ & $c$ & $\mu a$  \\ 
\hline
$V_{\rm NA}$   & 0.2368(1)  & -0.384(1)  & 0.8415(7)   \\ 
$V_{\rm A}$     & 0.21(5)     & -0.6(6)     & 2.7(4)  \\ 
\hline
\multicolumn{4}{l}{$\beta =5.6, 16^3\times 16$} \\ 
\hline
$V_{\rm NA}$  & 0.239(2)  & -0.39(4)  & 0.79(2)  \\ 
$V_{\rm A}$    & 0.25(2)   &  -0.3(1)   & 2.6(1)  \\
\hline
\multicolumn{4}{l}{$\beta =5.7, 12^3\times 12$} \\
\hline
$V_{\rm NA}$  & 0.159(3)  & -0.272(8)  & 0.79(1)  \\ 
$V_{\rm A}$    & 0.145(9)  & -0.32(2)   & 2.64(3) \\ 
\hline
\multicolumn{4}{l}{$\beta =5.8, 12^3\times 12$}\\ 
\hline
$V_{\rm NA}$  & 0.101(3) & -0.28(1)  &  0.82(1) \\ 
$V_{\rm A}$    & 0.102(9)  & -0.27(2)  &  2.60(3) \\
\hline
\end{tabular}
\end{center}
\end{table}
The best fitted values of the non-Abelian and Abelian string tensions are plotted in 
Table~\ref{stringtension_multilevel}.

\begin{figure}[h]
  \begin{minipage}[b]{0.5\linewidth}
    \centering
    \includegraphics[keepaspectratio, scale=0.35]{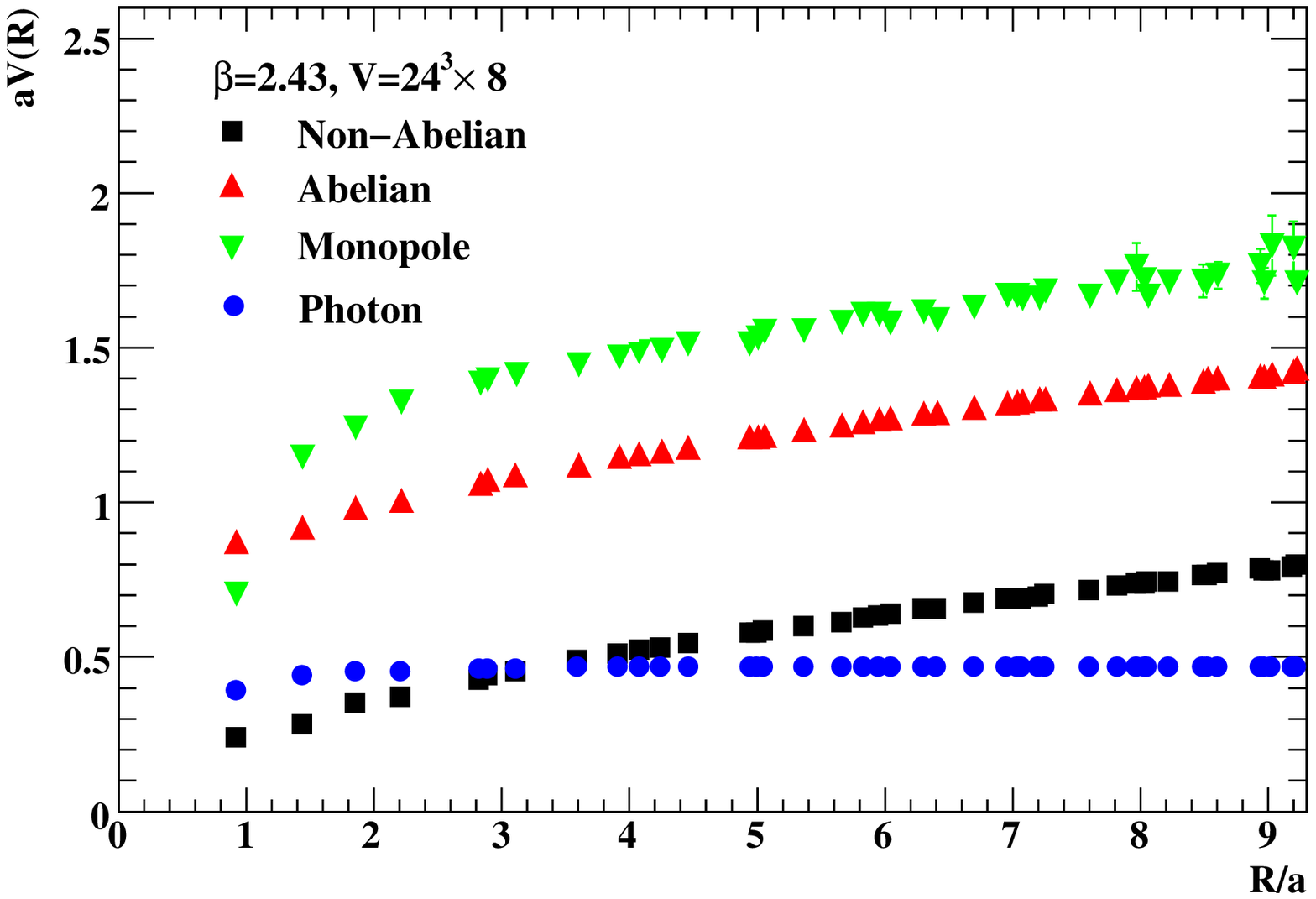}
    \subcaption{$SU(2)$ $24^3\times 8$ at $\beta=2.43$\cite{Suzuki:2009xy}}\label{FAMpotSU2}
  \end{minipage}
  \begin{minipage}[b]{0.5\linewidth}
    \centering
    \includegraphics[keepaspectratio, scale=0.35]{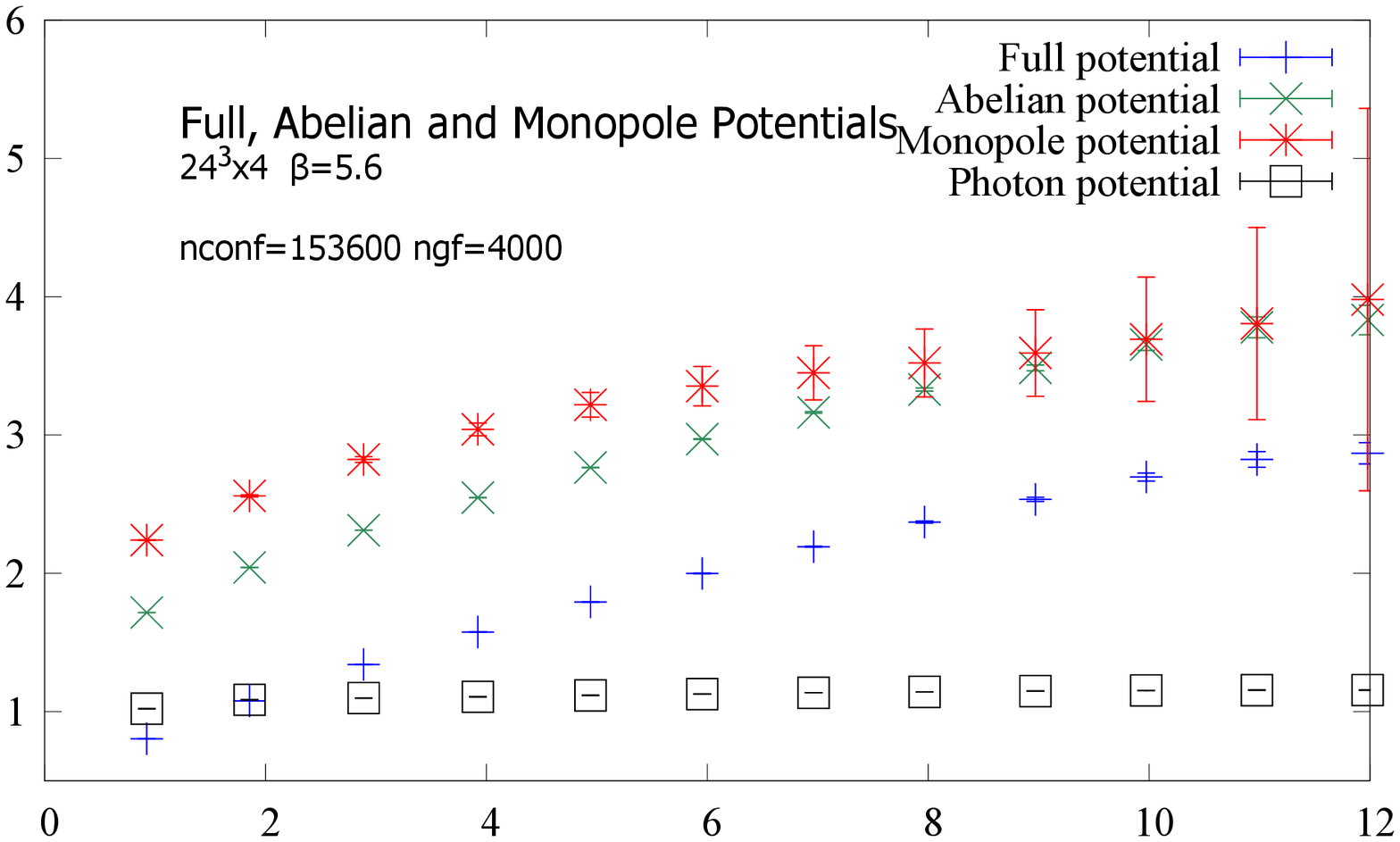}
    \subcaption{$SU(3)$ $24^3\times 4$ at $\beta=5.6$}\label{FAMpot_b56}
  \end{minipage}
  \caption{Non-Abelian, Abelian, monopole and photon static potentials} 
\end{figure}

\subsection{Monopole dominance in $SU(3)$}
Without adopting any further gauge fixing smoothing the vacuum, Abelian monopole static potential can reproduce fully the string tension of non-Abelian static potential in $SU(2)$ QCD as shown in Ref.~\cite{Suzuki:2007jp, Suzuki:2009xy}. This is called as perfect monopole dominance of the string tension. 

We investigate the monopole contribution to the $SU(3)$ static potential through Polyakov-loop correlators in order to examine the role of monopoles for confinement without any additional gauge-fixing. As shown below, it is much more difficult in $SU(3)$ than in $SU(2)$.  The monopole part of the Polyakov loop operator is extracted as follows.
Using the lattice Coulomb propagator $D(s-s')$, which satisfies
$\partial_{\nu}\partial'_{\nu}D(s-s') = -\delta_{ss'}$ with a
forward (backward) difference $\partial_{\nu}$ ($\partial'_{\nu}$), 
the temporal component of the Abelian fields $\theta^a_{4}(s)$ are written as 
\begin{equation}
\theta^a_4 (s) 
= -\sum_{s'} D(s-s')[\partial'_{\nu}\theta^a_{\nu 4}(s')+
\partial_4 (\partial'_{\nu}\theta^a_{\nu}(s'))] \; . 
\label{t4}
\end{equation} 
Inserting Eq.~\eqref{t4} to the Abelian Polyakov loop~\eqref{eq-PA},
we obtain
\begin{eqnarray}
&&P^a_{\rm A} = P^a_{\rm ph} \cdot P^a_{\rm mon}\; ,\nonumber\\
&&P^a_{\rm ph} = \exp\{-i\sum_{k=0}^{N_{t}-1} \!\sum_{s'}
D(s+k\hat4-s')\partial'_{\nu}\bar{\theta}^a_{\nu 4}(s')\} \; ,\nonumber\\
&&P^a_{\rm mon} = \exp\{-2\pi i\sum_{k=0}^{N_{t}-1}\! \sum_{s'}
D(s+k\hat4-s')\partial'_{\nu}n^a_{\nu 4}(s')\}\; .\nonumber\\
\label{ph-mon}
\end{eqnarray}
We call $P^a_{\rm ph}$ the photon and $P^a_{\rm mon}$ the monopole parts of 
the Abelian Polyakov loop $P_{\rm A}^a$, respectively~\cite{Suzuki:1994ay}.
The latter is due to the fact that the Dirac strings $n^a_{\nu 4}(s)$ lead to the monopole currents in 
Eq.~\eqref{eq:amon}~\cite{DeGrand:1980eq}.
Note that the second term of Eq.~\eqref{t4} does not contribute to the Abelian Polyakov loop 
in Eq.~\eqref{eq-PA}.
We show the simulation parameters and the results in comparison with the $SU(2)$ case.
\begin{table}[t]
\begin{center}
\caption{
Simulation parameters for the measurement of the static potential and the force from $P_{\rm A}$, $P_{\rm ph}$ and $P_{\rm mon}$. $N_{\rm RGT}$ is the number of random gauge transformations. The $SU(2)$ data are cited from Ref.\cite{Suzuki:2009xy}.}\label{SU3data}
\begin{tabular}{c|c|c|c|c|c}
\hline
&$\beta$ &$N_{s}^{3}\times N_{t}$& $a(\beta)$~[fm]& $N_{\rm conf}$ & $N_{\rm RGT}$ \\ 
\hline
$SU2$&2.43 &$24^{3} \times 8$ & 0.1029(4)& 7,000 & 4,000 \\
\hline
$SU3$&5.6 & $24^{3} \times 4$ & 0.2235\F & 153,600 & 4,000\\
\hline
\end{tabular}
\end{center}
\end{table}
 
\begin{table}
\centering
\fontsize{8pt}{20pt}\selectfont
\caption{
Best fitted values of the string tension $\sigma a^2$, the
Coulombic coefficient $c$, and the constant $\mu a$ for the
potentials $V_{\rm NA}$, $V_{\rm A}$, $V_{\rm mon}$ and $V_{\rm ph}$.
$V_{\rm FM}$ stands for the potential determined from non-Abelian and monopole  Polyakov loop correlators.}\label{FAMPFIT}
\begin{tabular}{l|l|c|c|c|c|c}
 \hline
$SU(2)$&&&&&&\\
$24^3\times 8$&$V_{\rm NA}$  & 0.0415(9)  & 0.47(2) & 0.46(8)  & 4.1 - 7.8\F & 0.99 \\ 
&$V_{\rm A}$ & 0.041(2)  & 0.47(6) & 1.10(3)  & 4.5 - 8.5\F & 1.00 \\ 
&$V_{\rm mon}$ & 0.043(3)  & 0.37(4) & 1.39(2)  & 2.1 - 7.5\F & 0.99 \\ 
&$V_{\rm ph}$ &$-6.0(3)\times 10^{-5}$ &0.0059(3) & 0.46649(6)  & 7.7 - 11.5 & 1.02\\
\hline
\hline
$SU(3)$&&&&&&\\
$24^3\times 4$&$V_{\rm NA}$ & 0.1707(47)  & 0.855(107) & 1.124(45)  & 3 - 10 & 1.170 \\ 
&$V_{\rm A}$ & 0.1771(84)   & 0.533(54)   & 3.56(6)  & 1 - 9 & 0.81 \\ 
&$V_{\rm FM}$ & 0.146(17)  & 0.345(29) & 2.477(59) & 0 - 8 & 0.970  \\ 
 \hline
\end{tabular}
\end{table}  
For comparison, we plot the $SU(3)$ data as well as a typical old $SU(2)$ data~\cite{Suzuki:2007jp, Suzuki:2009xy} . In the $SU(2)$ case,  beautiful Abelian and monopole dominances are observed using reasonable number of vacuum ensembles as seen from Fig.\ref{FAMpotSU2} and Table\ref{SU3data}. In $SU(3)$, we needed much more number of vacuum ensembles even on $24^3\times 4$ small lattice in $SU(3)$ as shown in Tabel~\ref{SU3data}. Abelian dominance is seen from Abelian-Abelian Polyakov loop correlators. But in the case of monopole-monopole Polyakov loop correlators, we could not get good results. Hence we try to study non-Abelian and monopole correlators $V_{FM}$.  Since the fit is not good enough, monopole dominance is seen from the hybrid correlators as shown in Fig.\ref{FAMpot_b56} and Table\ref{FAMPFIT}.
 To improve monopole static potentials for the large separation region, we may study larger $T$ lattice with much more vacuum ensembles.
 In the case of photon-photon correlators, the string tensions on both cases are almost zero.
 \section{The Abelian dual Meissner effect in $SU(3)$}
\subsection{Simulation details of the flux-tube profile}
In this section, we show the results with respect to the Abelian dual Meissner effect. In the previous work \cite{Suzuki:2009xy} studying the spatial distribution of color electric fields and monopole currents, they  used the connected correlations between a non-Abelian Wilson loop and Abelian operators in $SU(2)$ gauge theory without gauge fixing. We apply the same method to $SU(3)$ gauge theory without gauge fixing. Here we employ the standard Wilson action on the $24^3(40^3) \times 4$ lattice with the coupling constant $\beta = 5.6$. We consider a finite temperature system at $T = 0.8 T_c$. To improve the signal-to-noise ratio, the APE smearing is applied to the spatial links and the hypercubic blocking is applied to the temporal links.  We introduce  random gauge transformations to improve the signal to noise ratios of the data concerning the Abelian operators. 

\begin{figure}[h]
\begin{tabular}{cc}
  \begin{minipage}[b]{0.5\linewidth}
    \centering
\includegraphics[keepaspectratio, scale=0.6]{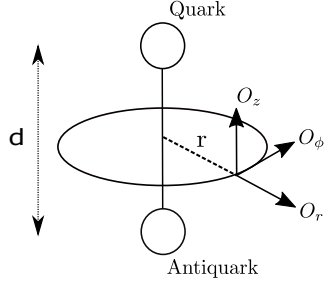}
\subcaption{The cylindrical coordinate}\label{cor}  
  \end{minipage}&
\begin{minipage}[b]{0.5\linewidth}
     \centering
    \includegraphics[keepaspectratio, scale=0.45]{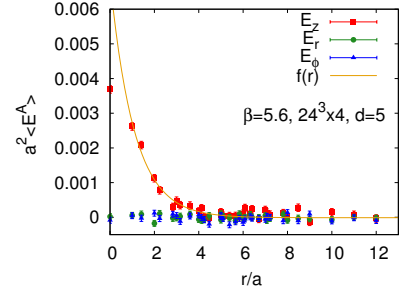}  

     
\subcaption{The Abelian color electric field} \label{EA_zrp}   
  \end{minipage}
  \end{tabular}
  \caption{Profile of the coordinate and the Abelian color electirc field around static quarks for $d=5$ at $\beta = 5.6$ on $24^3\times 4$ lattices.}
\end{figure}

To measure the flux-tube profiles, we consider a connected correlation functions as done in \cite{Cea2016, refId0}:  
  \begin{align}
  \rho_{conn}(O(r)) = &\frac{\kbra{\Tr(P(0)LO(r)L^{\dagger})\Tr P^{\dagger}(d)}}{\kbra{\Tr P(0) \Tr P^{\dagger}(d)}} 
   - \frac{1}{3}\frac{\kbra{\Tr P(0) \Tr P^{\dagger}(d) \Tr O(r)}}{\kbra{\Tr P(0) \Tr P^{\dagger}(d)}}, \label{connect}
 \end{align} 
where $P$ denotes a non-Abelian Polyakov loop, $L$ indicates a Schwinger line, $r$ is a distance from a flux-tube and $d$ is a distance between Polyakov loops. We use the cylindrical coordinate $(r,\phi,z)$ to parametrize the $q\text{-}\bar{q}$ system as shown in Fig.\ref{cor}. Here  the definition of the cylindrical coordinate $(r, \phi, z)$ along the $q\text{-}\bar{q}$ axis. The $d$ corresponds to the distance between Polyakov loops.

\subsection{The spatial distribution of color electric fields}
First of all, we show the results of Abelian color electric fields using an Abelian gauge field $\theta_1(s,\mu)$. To evaluate the Abelian color electric field, we adopt the Abelian plaquette as an operator $O(r)$. We calculate a penetration length $\lambda$ from the Abelian color electric fields for $d=3,4,5,6$ at $\beta=5.6$ and check the $d$ dependence of $\lambda$. To improve the accuracy of the fitting, we evaluate  $O(r)$ at both on-axis and off-axis distances. As a result, we find the Abelian color electric fields $E^A_{z}$ alone are squeezed as in Fig.\ref{EA_zrp}.  We fit these results to a fitting function, 
  \begin{align}
   f(r) = c_1 \mathrm{exp}(-r/\lambda) + c_0 .
  \end{align}  
The parameter $\lambda$ corresponds to the penetration length. We find the values of the penetration length are almost the same.

\subsection{The spatial distribution of monopole currents}
Next we show the result of the spatial distribution of Abelian-like monopole currents.  We define the Abelian-like monopole currents on the lattice as in Eq.~(\ref{eq:amon}).
In this study we evaluate the connected correlation (\ref{connect}) between $k^1$ and the two non-Abelian Polayakov loops. We use random gauge transformations to evaluate this correlation. As a result, we find the spatial distribution of monopole currents around the flux-tube at $\beta = 5.6$. Only the monopole current in the azimuthal direction, $k^1_{\phi}$, shows the correlation with the two non-Abelian Polyakov loops as shown in Fig.\ref{mono}.

\subsubsection{The dual Amp\`{e}re's law}
In previous $SU(2)$ researches \cite{Suzuki:2009xy}, they investigated the dual Amp\`{e}re's law to see what squeezes the color-electric field. In the case of $SU(2)$ gauge theory without gauge fixings, they confirmed the dual Amp\`{e}re's law and the monopole currents squeeze the color-electric fields. In this section we show the results of the dual Amp\`{e}re's law in the case of $SU(3)$ gauge theory. The definition of monopole currents gives us  the following relation,
\begin{align}
 (\mathrm{rot}E^{a})_\phi = \partial_{t}B^{a}_{\phi} + 2\pi k^{a}_{\phi}, 
\end{align}
where index $a$ is a color index. 

\begin{figure}[h]
  \begin{minipage}[b]{0.5\linewidth}
    \centering
    \includegraphics[keepaspectratio, scale=0.5]{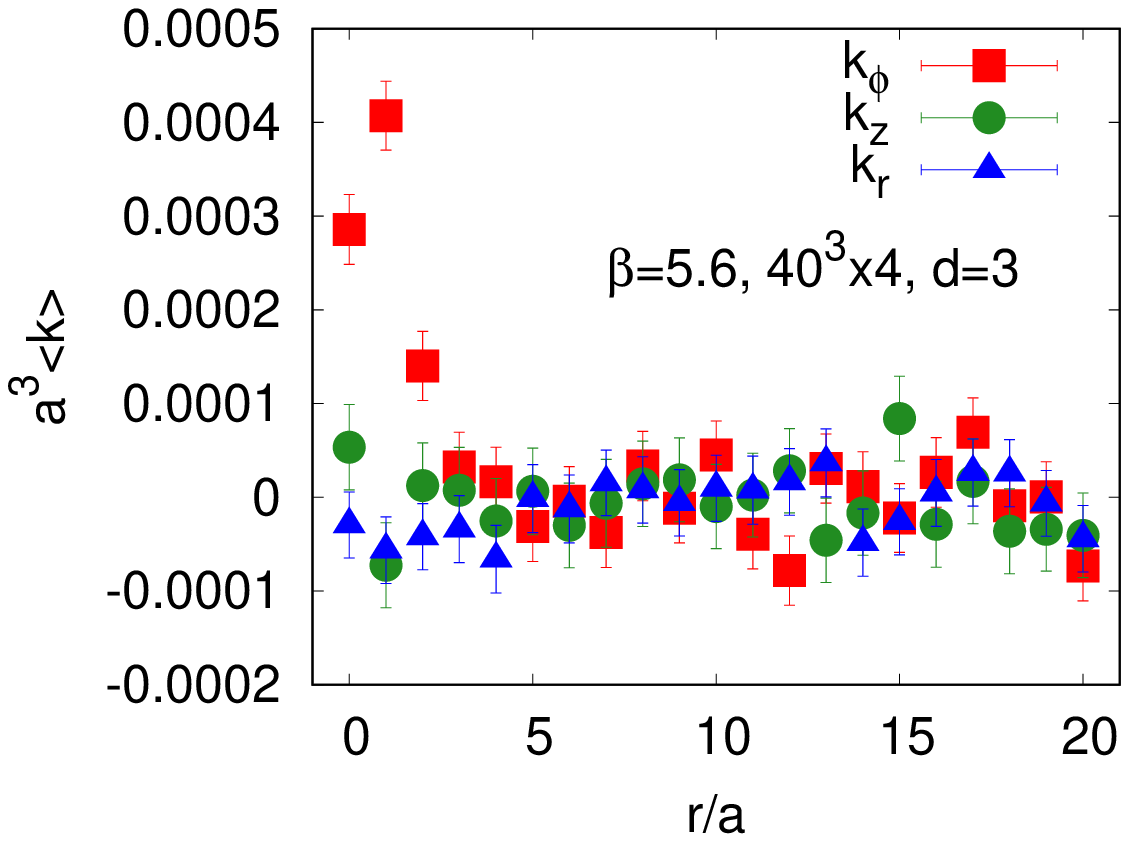}
  
\subcaption{The profile of monopole current $ k_{\phi}, k_z, k_r$}\label{mono}        
  \end{minipage}
  \begin{minipage}[b]{0.5\linewidth}
\centering
\includegraphics[keepaspectratio, scale=0.5]{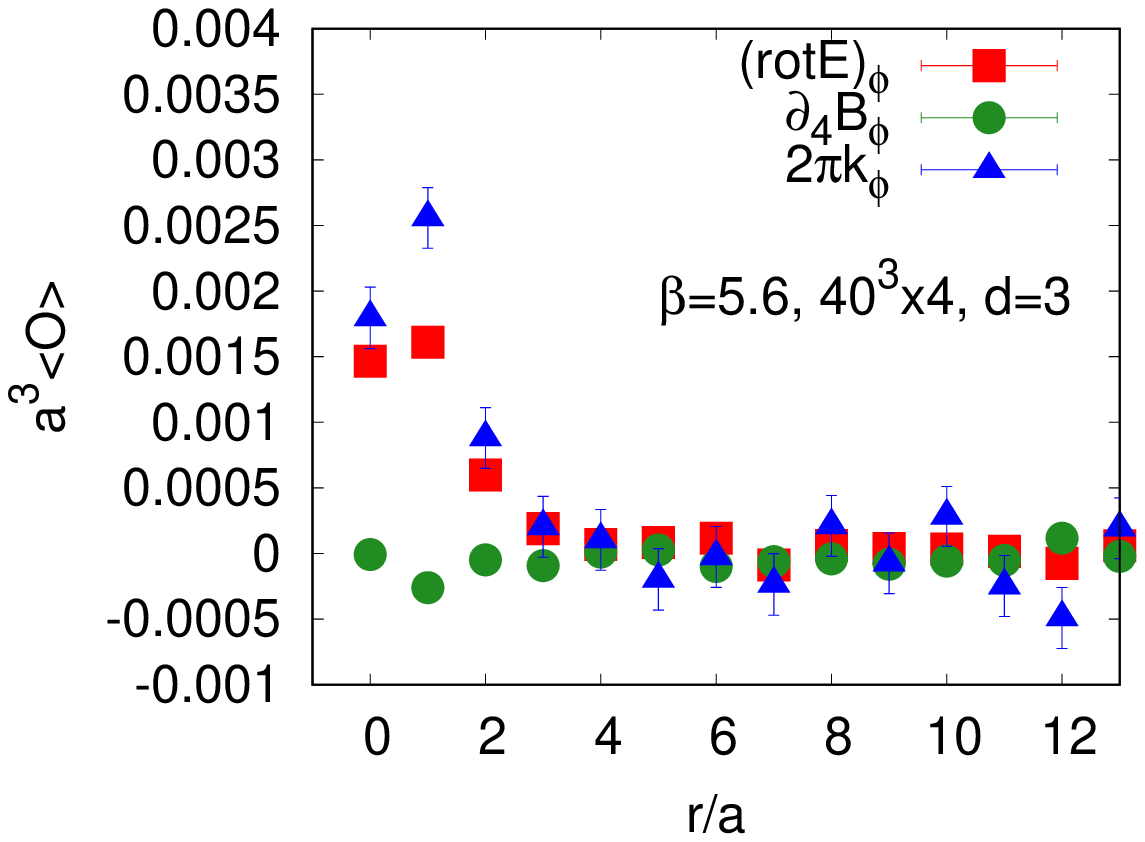}

\subcaption{The dual Amp\`{e}re's law} \label{dualA}
  \end{minipage}
\caption{The monopole current on $40^3\times 4$ at $\beta = 5.6$.}
\end{figure}

As a results, we confirm that there is no signal of the magnetic displacement current $\partial_{t}B^{a}_{\phi}$ around the flux-tube for $d=3$ at $\beta=5.6$ as shown in Fig.\ref{dualA}. It suggests that the Abelian-like monopole current squeezes the Abelian color electric field as a solenoidal current in $SU(3)$ gauge theory without gauge fixing, although more data for larger $d$ are necessary.
 
\subsection{The vacuum type in $SU(3)$ gauge theory without gauge fixing}
Finally, we evaluate the Ginzburg-Landau parameter, which characterizes the type of the (dual) superconducting vacuum. In the previous result \cite{Suzuki:2009xy}, they found that the vacuum type is near the border between type 1 and type 2 dual superconductors by using the $SU(2)$ gauge theory without gauge fixing. We apply the same method to $SU(3)$ gauge theory. 

\begin{figure}[h]
\begin{center}
\includegraphics[width=7cm]{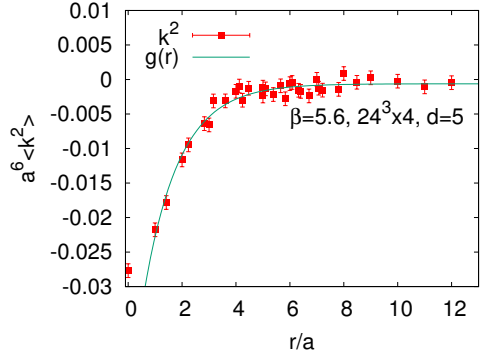}
\end{center}
\caption{The squared monopole density with $d=5$ at $\beta = 5.6$ on $24^3\times 4$ lattices.}\label{kk}
\end{figure}

To evaluate the coherence length, we measure the correlation between the squared monopole density and two non-Abelian Polyakov loops by using the disconnected correlation function~\cite{PhysRevD.72.074505,Suzuki:2009xy},
\begin{align}
 \kbra{k^2(r)}_{q\bar{q}} = &\frac{\kbra{\Tr{P(0)}\Tr{P^{\dagger}(d)}\sum_{\mu, a}k^a_{\mu}(r)k^a_{\mu}(r)}}{\kbra{\Tr{P}(0)\Tr{P^{\dagger}(d)}}} 
 - \kbra{\sum_{\mu,a}k^a_{\mu}(r)k^a_{\mu}(r)}.
\end{align}
We fit the profiles to the function,
\begin{align}
 g(r) = c'_1 \mathrm{exp}\left(-\frac{\sqrt{2}r}{\xi}\right) + c'_0,
\end{align}
where the parameter $\xi$ corresponds to the coherence length. We plot the profiles of $\kbra{k^2(r)}_{q\bar{q}} $ in Fig.\ref{kk}.  As a result, we could evaluate the coherence length $\xi$ for $d=3,4,5,6$ at $\beta=5.6$ and find almost the same values of $\xi/\sqrt{2}$ for each $d$. Using these parameters $\lambda$ and $\xi$, we could evaluate the Ginzburg-Landau parameter. The GL parameter  $\kappa = \lambda/\xi$ can be defined as the ratio of the penetration length and the coherence length. If $\sqrt{2}\kappa < 1$, the vacuum type is of the type 1 and if $\sqrt{2}\kappa > 1$, the vacuum  is of the type 2.  We show the GL parameters in $SU(3)$ gauge theory in Table \ref{GL}. We find that the vacuum  is of the type 1 near the border between type 1 and type 2, although the study is done at one gauge coupling constant $\beta=5.6$. This is the first result of the vacuum type in pure $SU(3)$ gauge theory without gauge fixing, although different $\beta$ data are necessary to show the continuum limit. 
\begin{table}[h] 
\begin{center}
\caption{The Ginzburg-Landau parameters at $\beta=5.6$ on $24^3\times 4$ lattice.}\label{GL}
\begin{tabular}{|c|c|} \hline
 \multicolumn{1}{|c|}{d} &  \multicolumn{1}{|c|}{$\sqrt{2}\kappa$} \\ \hline
3&0.87(5) \\
\hline
4&0.93(7)  \\
\hline
5&0.83(9) \\
\hline
6&0.9(2)  \\
\hline
\end{tabular}
\end{center}
\end{table}

\section*{Acknowledgements}
The authors would like to thank Y. Koma for his computer  code of the 
multilevel method. The numerical simulations of this work were done using High Performance Computing resources at Cybermedia Center and Research Center for Nuclear Physics  of Osaka University,  at Cyberscience Center of Tohoku University and at KEK. The authors would like to thank these centers for their support of computer facilities. T.S  was finacially supported by JSPS KAKENHI Grant Number JP19K03848.




\nolinenumbers


\begin{thebibliography}{88}
\bibitem{tHooft:1975pu}
G.~'t~Hooft,
\newblock in {\em Proceedings of the EPS International}, edited by A.~Zichichi,
  p. 1225, 1976.
 
\bibitem{Mandelstam:1974pi}
S.~Mandelstam,
\newblock Phys. Rept. {\bf 23}, 245 (1976).

\bibitem{Dirac:1931} 
 P. Dirac,  Proc. Roy. Soc. (London) A 133, 60 (1931).

\bibitem{tHooft:1981ht}
G.~'t~Hooft,
\newblock Nucl. Phys. {\bf B190}, 455 (1981).


\bibitem{Kronfeld:1987ri}
A.~S. Kronfeld, M.~L. Laursen, G.~Schierholz, and U.~J. Wiese,
\newblock Phys. Lett. {\bf B198}, 516 (1987).

\bibitem{Kronfeld:1987vd}
A.~S. Kronfeld, G.~Schierholz, and U.~J. Wiese,
\newblock Nucl. Phys. {\bf B293}, 461 (1987).

\bibitem{Suzuki:1992rw}
T.~Suzuki,
\newblock Nucl. Phys. Proc. Suppl. {\bf 30}, 176 (1993).

\bibitem{Chernodub:1997ay}
M.~N. Chernodub and M.~I. Polikarpov,
\newblock 
\newblock in {\em "Confinement, Duality and Nonperturbative Aspects of QCD"},
  edited by P.~van Baal, p. 387, Cambridge, 1997, Plenum Press.


\bibitem{Bonati:2010tz}
C.~Bonati, A.~Di~Giacomo, L.~Lepori and F.~Pucci, 
\newblock Phys. Rev. {\bf D81}, 085022 (2010).

\bibitem{Suzuki:2014wya}
Tsuneo Suzuki, A new scheme for color confinement due to violation of
the non-Abelian Bianchi identities, arXiv:1402.1294  

\bibitem{Arafune:1974uy}
J.~Arafune,  P.G.O.~Freund  and C.J.~Goebel, 
\newblock J.Math.Phys. {\bf 16}, 433 (1975).


\bibitem{Luscher:2001}
M.L\"{u}scher and P.Weisz, JHEP {\bf 09}, 010 (2001)


\bibitem{Suzuki:2007jp}
T.~Suzuki, K.~Ishiguro, Y.~Koma and T.~Sekido,
\newblock Phys. Rev. {\bf D77}, 034502 (2008).

\bibitem{Suzuki:2009xy}
T.~Suzuki, M.~Hasegawa, K.~Ishiguro, Y.~Koma and T.~Sekido,
\newblock Phys. Rev. {\bf D80}, 054504 (2009).


\bibitem{Suzuki:2017lco}
Tsuneo Suzuki, Katsuya Ishiguro and Vitaly Bornyakov,  Phys. Rev. {\bf D97}, 034501 (2018). Erratum: Phys. Rev. {\bf D97}, 099905 (2018).

\bibitem{Suzuki:2017zdh}
Tsuneo Suzuki, Phys. Rev.  {\bf D 97} 034509 (2018). 

\bibitem{DeGrand:1980eq}
T.~A. DeGrand and D.~Toussaint,
\newblock Phys. Rev. {\bf D22}, 2478 (1980).


\bibitem{Suzuki:1994ay}
T.~Suzuki et al., Phys. Lett. {bf B347}, 375 (1995).

\bibitem{DelDebbio:1996mh}
L. ~Del Debbio, M. ~Faber, J. ~Greensite and S. ~Olejnik,
\newblock	Phys. Rev. \textbf{D55}, 2298 (1997)
\bibitem{DelDebbio:1998uu}
L. ~Del Debbio, M. ~Faber, J. ~Giedt, J. ~Greensite and S. ~Olejnik,
\newblock	Phys. Rev. \textbf{D58}, 094501 (1998)

\bibitem{Faber:2001zs}
M. ~Faber, J. ~Greensite and S. ~Olejnik,
\newblock JHEP {\bf 111}, 053 (2001).

\bibitem{Suzuki:1996ax}
T. ~Suzuki et al.,
\newblock Nucl. Phys. Proc. Suppl. {\bf 53}, 531 (1997).
\bibitem{Cea2016}
Paolo Cea, Leonardo Cosmai, Francesca Cuteri, and Alessandro Papa,
Journal of High Energy Physics, 2016, 6 (2016).

\bibitem{refId0}
Paolo Cea, Leonardo Cosmai, Francesca Cuteri, and Alessandro Papa,
EPJ Web Conf., 175, 12006 (2018).

\bibitem{PhysRevD.72.074505}
M.~N. Chernodub, Katsuya Ishiguro, Yoshihiro Mori, Yoshifumi Nakamura, M.~I.
  Polikarpov, Toru Sekido, Tsuneo Suzuki, and V.~I. Zakharov,
Phys. Rev. {\bf D72}, 074505 (2005).


\end{thebibliography}
\end{document}